\documentclass[twocolumn]{article}
\usepackage[a4paper, total={7in, 9.5in}]{geometry}

\usepackage[T1]{fontenc}
\usepackage[utf8]{inputenc}
\usepackage{times}
\usepackage{amsmath,amsfonts}
\DeclareMathOperator*{\argmin}{arg\,min}
\usepackage{graphicx}
\usepackage{algorithm}
\usepackage{algpseudocode}
\usepackage{booktabs}
\usepackage{titlesec}
\usepackage{caption}
\usepackage{fancyhdr}
\usepackage{multicol}
\usepackage{url}
\usepackage{abstract}
\usepackage{multirow} 
\title{\textbf{Cross-platform epistemic verification for improving factual reliability in AI-generated news summarization}}

\author{
    Zhuo Xie$^{1}$, Haoze Ni$^{2}$\\
    {\small $^{1}$Bank of Changsha Co., Ltd., China}\\
    {\small $^{2}$College of Communication, Emerging Media Studies, Boston University, Boston, United States}\\
    {\small Corresponding author: zhuoxie.uon@gmail.com}
}
\date{}

\begin{document}

\maketitle

\begin{abstract}

Large Language Models (LLMs) are increasingly used in automated news summarization and digital content production. However, generated summaries may contain factual errors or hallucinated information, raising concerns about the reliability of AI-generated news content. Existing post-hoc correction methods typically rely on a single retrieval source for verification, making them sensitive to retrieval bias, incomplete evidence, and platform-specific inconsistencies.

This study proposes Multi-source Evidence Consensus Verification (MECV), a post-hoc hallucination correction framework for AI-generated news summarization. Instead of depending on a single retrieval channel, MECV aggregates evidence from multiple heterogeneous sources, including the source document, Wikipedia, and open-web retrieval. The framework further incorporates a multi-LLM jury mechanism that estimates factual reliability through contradiction-aware consensus scoring across verifier models. Claims identified as potentially unsupported are revised through iterative minimal-edit refinement.

The proposed framework is evaluated on the SummEdits benchmark using GPT-4o-mini and DeepSeek-Chat as the verifier jury, with Qwen-Plus as the orchestrator. Experimental results show that MECV improves factual consistency while preserving the semantic structure of the original summaries. The findings further suggest that agreement across heterogeneous evidence sources can serve as a useful signal for identifying factual uncertainty in AI-generated summaries, including in information-sensitive domains such as financial news aggregation.

This study contributes to research on trustworthy AI and automated journalism by introducing a multi-source verification framework for hallucination correction and demonstrating the value of consensus-based verification for improving factual reliability in AI-generated news summarization.

\end{abstract}

\vspace{1em}

\noindent\textbf{Keywords:}
Large Language Models,
Hallucination Correction,
Factual Consistency,
Post-hoc Verification,
Multi-Engine Retrieval,
Consensus Verification,
News Summarization,
Retrieval-Augmented Generation,
Financial News Summarization,
Fact Verification,
LLM Self-Correction
\section{Introduction}
The growing presence of generative artificial intelligence in digital news environments has created both opportunities and challenges for online information systems. Large Language Models (LLMs) are increasingly integrated into journalistic workflows, supporting tasks such as automated summarization, multilingual content production, recommendation, and editorial assistance \cite{pavlik2023collaborating,caswell2024audiences}. These applications have expanded the capacity of news platforms to process and distribute information at scale. At the same time, the wider adoption of AI-assisted content production has raised broader questions about information quality, credibility, and public trust in digitally mediated communication.

Among these concerns, factual reliability has emerged as a central issue. Although LLMs are capable of producing coherent and highly fluent language, they may also generate content that is inaccurate, fabricated, or unsupported by evidence--a phenomenon commonly referred to as hallucination \cite{ji2023survey}. In news summarization tasks, hallucinations may appear as substituted entities, invented quotations, distorted causal relationships, or temporal inconsistencies. Such errors carry consequences beyond technical performance. AI-generated summaries and recommendation interfaces are increasingly becoming part of how audiences encounter and navigate news content online. This is particularly important for time-sensitive domains such as financial news, where summaries of earnings announcements, macroeconomic releases, regulatory actions, or market events may be used as inputs to investor interpretation and downstream information services. As a result, factual distortions embedded within generated summaries may shape audience understanding and influence perceptions of social, political, and economic events \cite{sundar2008main}. Improving factual consistency has therefore become an important requirement for trustworthy AI-mediated communication.

A growing body of research has explored methods for reducing hallucinations in generated text, including retrieval-augmented generation, model self-correction, reinforcement learning approaches, and post-hoc verification frameworks \cite{gao2023retrieval,dhuliawala2024chain}. Among these approaches, post-hoc correction has received increasing attention because it can improve factual reliability without modifying the underlying language model. Typical systems extract factual claims from generated summaries, retrieve external evidence, and revise unsupported content through iterative correction procedures.

Despite these advances, many existing verification systems continue to rely on a single retrieval source to determine factual validity. This design assumption has received relatively limited attention, despite its implications for information quality. Different retrieval channels--including search engines, encyclopedic resources, and news databases--vary considerably in indexing coverage, update frequency, and source prioritization. Consequently, factual judgments may become dependent on the characteristics of a particular retrieval system rather than on the underlying truth status of a claim. This limitation is especially important in contemporary online information environments, where information is fragmented across heterogeneous and rapidly evolving sources.

More importantly, this reliance on isolated evidence differs from long-standing journalistic verification practices. Professional fact-checking traditionally depends on corroboration across independent sources rather than validation through a single informational authority \cite{kovach2007elements,ward2015strengthening}. Claims receiving inconsistent support across multiple channels are generally treated with greater caution. Yet existing AI-based hallucination correction systems have paid relatively limited attention to evidence diversity and cross-source agreement as signals of factual uncertainty.

To address this gap, this study introduces Multi-source Evidence Consensus Verification (MECV), a post-hoc hallucination correction framework for AI-generated news summarization. MECV combines evidence from heterogeneous information sources--including source documents, Wikipedia, and open-web retrieval--and employs a consensus-based verification process to identify potentially unsupported claims. Suspected hallucinations are then revised through a conservative minimal-edit correction strategy designed to preserve semantic fidelity wherever possible.

The framework is evaluated using the SummEdits benchmark \cite{laban2023summedits} with GPT-4o-mini and DeepSeek-Chat as the verifier jury and Qwen-Plus as the orchestrator. In addition to automatic evaluation metrics, human assessments involving annotators with journalism and information science backgrounds were conducted to examine perceived credibility and readability. Results indicate that multi-source consensus verification improves factual consistency while reducing unnecessary rewriting behaviors commonly observed in conventional retrieval-based correction systems.

This study makes three contributions. First, it introduces a multi-source post-hoc verification framework grounded in principles of journalistic corroboration. Second, it demonstrates that cross-source agreement can function as a practical signal of factual uncertainty and improve factual reliability while mitigating over-correction. Third, it provides empirical evidence on the role of evidence diversity in improving the trustworthiness of AI-generated news summaries.

More broadly, the findings suggest that future AI-assisted information systems may benefit from verification mechanisms inspired by long-established journalistic practices of corroboration and source diversity. As generative AI becomes increasingly embedded within digital communication infrastructures, improving the credibility of AI-generated information may depend not only on stronger language models, but also on more robust approaches to validating information across complex online environments.

\section{Related Work}

We review related work from five perspectives: hallucination and factuality in abstractive news summarization, factual consistency evaluation, post-hoc hallucination correction, retrieval-augmented verification, and consensus-based verification with large language models. These directions jointly motivate our framework, which treats hallucination correction not only as a text revision problem, but also as an evidence-grounded consensus process across heterogeneous information sources and model judges.

\subsection{Hallucination and Factuality in News Summarization}

Factual inconsistency has been widely recognized as a central limitation of abstractive summarization. Although neural summarization models can generate fluent and coherent summaries, they may introduce entities, relations, temporal expressions, or quantities that are not supported by the source document. Maynez et al.~\cite{maynez2020faithfulness} showed that abstractive summarizers frequently produce unsupported content and that factuality may conflict with abstractiveness. Kryscinski et al.~\cite{kryscinski2020evaluating} further argued that traditional lexical-overlap metrics are insufficient for evaluating factual consistency, since summaries can appear similar to references while still contradicting the source.

Later studies developed more fine-grained analyses of factual errors. Pagnoni et al.~\cite{pagnoni2021frank} introduced FRANK, a benchmark and taxonomy covering entity, predicate, circumstance, and discourse-level errors in abstractive summaries. Ji et al.~\cite{ji2023survey} provided a broader survey of hallucination in natural language generation and identified insufficient grounding and over-reliance on parametric memory as two recurring causes. For news summarization, Laban et al.~\cite{laban2023summedits} proposed SummEdits, where factual and hallucinated summaries differ only through minimal lexical edits. This setting is particularly challenging because the generated summary may remain fluent and semantically close to the original while containing a subtle factual distortion.

These studies demonstrate that hallucination in summarization is not merely a surface-level generation error. In news-oriented applications, even a small factual change can misrepresent events, actors, or outcomes. Therefore, hallucination correction for news summarization requires fine-grained claim verification rather than only global similarity measurement.

\subsection{Factual Consistency Evaluation and Claim-level Verification}

A substantial body of work has focused on automatic factuality evaluation. Early methods often formulate factual consistency as a classification or entailment problem. FactCC~\cite{kryscinski2020evaluating} trains a weakly supervised model to determine whether a summary is factually consistent with its source document. DAE~\cite{goyal2020evaluating} evaluates factuality through dependency-level entailment, allowing factual errors to be localized at a more fine-grained semantic level.

Question-answering-based methods provide another important evaluation paradigm. FEQA~\cite{durmus2020feqa} generates questions from a summary and checks whether the answers can be recovered from the source document. QAGS~\cite{wang2020qags} similarly compares answers derived from the summary and source to estimate factual consistency. These methods move beyond lexical overlap and attempt to test whether the information expressed in the summary is actually supported by the source.

With the emergence of large language models, claim-level factuality evaluation has become increasingly common. FActScore~\cite{min2023factscore} decomposes generated text into atomic factual claims and computes the proportion of claims supported by external evidence. G-Eval~\cite{liu2023geval} uses large language models with explicit evaluation rubrics and shows stronger alignment with human judgments in natural language generation evaluation. These methods provide useful tools for assessing factual reliability, but they primarily focus on evaluation rather than producing a corrected output. In contrast, our work uses atomic claim verification as part of an iterative correction pipeline.

\subsection{Post-hoc Hallucination Correction}

Post-hoc correction has become a practical solution for improving factual reliability when the underlying LLM cannot be retrained. Instead of changing model parameters, post-hoc systems verify and revise the generated output after generation. RARR~\cite{gao2023rarr} retrieves external evidence for model-generated claims and revises unsupported statements through minimal edits. This approach is particularly relevant to closed-source LLMs because it can be applied as an external correction layer.

Self-verification and self-refinement methods represent another line of work. Chain-of-Verification~\cite{dhuliawala2024chain} asks an LLM to generate verification questions, answer them independently, and revise its initial response. Self-Refine~\cite{madaan2023selfrefine} generalizes iterative feedback and revision across different generation tasks. More recently, Vladika et al.~\cite{vladika2025correcting} explored self-correcting LLM methods with external knowledge for hallucination correction in news summaries, showing the relevance of external evidence for news-oriented factual correction.

Despite their effectiveness, existing post-hoc methods have two important limitations. First, self-correction methods may rely on the same model for both generation and verification, which can reproduce the original reasoning bias. Second, retrieval-augmented correction methods often rely on a single retrieval channel, making the revision process sensitive to retrieval bias, incomplete evidence, and source coverage limitations. These limitations motivate a correction framework that combines external evidence with heterogeneous verification signals.

\subsection{Retrieval-Augmented Verification and Evidence Reliability}

Retrieval-augmented generation has been widely used to improve the factual grounding of language models. RAG~\cite{lewis2020retrieval} introduced a general framework that combines parametric generation with non-parametric retrieval, enabling models to access external knowledge during generation. Retrieval-augmented language model pre-training methods such as REALM~\cite{guu2020realm} further demonstrate the value of integrating retrieval into language modeling. Recent surveys also emphasize that retrieval quality, evidence relevance, and source reliability remain critical challenges for RAG systems~\cite{zhao2024retrieval}.

The issue of evidence reliability is closely related to fact verification. FEVER~\cite{thorne2018fever} formalized fact verification as retrieving evidence and predicting whether a claim is supported, refuted, or unverifiable. However, many fact verification systems assume a relatively fixed evidence corpus. In open-web settings, evidence is more dynamic and heterogeneous. Search engines differ in indexing coverage, ranking mechanisms, freshness, and source prioritization. As a result, factual verification based on a single search engine may produce unstable or incomplete evidence.

For AI-generated news summaries, this problem is especially important. Journalistic verification traditionally relies on corroboration across multiple sources rather than a single authority. Therefore, hallucination correction for news summarization should not treat retrieved evidence as automatically reliable. Instead, it should model agreement and disagreement across evidence sources. Our framework follows this intuition by using multi-source evidence aggregation and contradiction-aware voting to estimate the reliability of each atomic claim.

\subsection{LLM-as-a-Judge, Multi-agent Verification, and Consensus Reasoning}

Large language models have also been increasingly used as evaluators and judges. Zheng et al.~\cite{zheng2023judging} studied the LLM-as-a-judge paradigm with MT-Bench and Chatbot Arena, showing that strong LLM judges can approximate human preference judgments while still exhibiting systematic biases. This suggests that a single LLM judge may not be sufficiently reliable for high-stakes factual verification, especially when the judge shares similar assumptions or knowledge limitations with the generator.

Multi-agent and ensemble-based methods attempt to reduce single-model bias through aggregation. Du et al.~\cite{du2023debate} showed that multiple LLMs debating a proposition can improve factuality and reasoning compared with individual model responses. Wang et al.~\cite{wang2024moa} proposed Mixture-of-Agents, demonstrating that aggregating outputs from multiple LLMs can improve overall model capability. These studies suggest that consensus among heterogeneous models can provide a useful signal for more robust reasoning and judgment.

However, most multi-agent approaches focus on open-ended reasoning, preference evaluation, or answer generation. Their application to atomic, evidence-grounded hallucination correction in news summarization remains underexplored. Moreover, prior work often aggregates model opinions without explicitly considering disagreement among retrieval sources. Our work addresses this gap by combining atomic claim decomposition, multi-source evidence aggregation, and multi-LLM contradiction voting. In this way, hallucination correction is formulated as a consensus process over both evidence sources and model judgments.

Taken together, prior work has established the importance of factuality evaluation, retrieval-augmented correction, and multi-model verification. However, existing systems often depend on a single verifier, a single retriever, or a single evidence source. This assumption is problematic for AI-generated news summarization, where factual reliability depends on source credibility, evidence coverage, and cross-source corroboration. MECV addresses this gap by modeling hallucination correction as an epistemic consensus process across heterogeneous evidence channels and LLM jurors.

\subsection{Information Quality and Credibility in AI-Assisted News Summarization}

In AI-assisted news summarization, factual consistency is part of a broader information quality problem. A generated summary is useful only when it provides information that is accurate, complete, consistent, and suitable for the user's information needs \cite{wang1996dataquality}. This is especially important in news settings, where summaries may serve as the first point of contact between readers and a news event. The same requirement applies to financial information services, where short summaries of market-moving events are often consumed before users inspect the underlying source material. In such settings, even small errors in entities, quantities, dates, or causal relations can reduce the decision value of the summary.

Credibility is also central to automated news production. Prior research on computer-generated news has shown that readers evaluate automated articles not only by factual accuracy, but also by perceived credibility, readability, and journalistic quality \cite{graefe2018computergenerated}. Therefore, hallucination correction should not be treated only as a technical process of replacing unsupported claims. It also needs to preserve the communicative function of the news summary.

A related concern is framing. News summaries do not merely list facts; they also organize information by selecting what is emphasized and how events are presented \cite{entman1993framing}. If a correction system rewrites too aggressively, it may improve factual support while unintentionally changing the original emphasis or interpretation of the report. In financial news, for example, changing whether a result is framed as an earnings surprise, a regulatory risk, or a temporary market fluctuation can alter the informational meaning of a summary even when the individual facts appear supported. For this reason, a reliable correction framework should improve factual consistency while avoiding unnecessary changes to tone, structure, and framing. This consideration is consistent with the minimal-edit design of MECV, which aims to correct unsupported claims without broadly rewriting the summary.

\section{Methodology}
\label{sec:method}

We propose MECV (Multi-source Evidence Consensus Verification), a
post-hoc framework that detects and revises hallucinations in news
summaries through three stages: claim extraction, contradiction
scoring by a heterogeneous LLM jury over multi-source evidence, and
RARR-style iterative minimal-edit refinement. The design is motivated
by two failure modes that single-source or single-verifier post-hoc
systems commonly face. First, retrieval-based verifiers that depend
on a single external search backend inherit its indexing, freshness,
and ranking biases; a fact may be \emph{retrievably} unsupported from
one source even when other evidence channels support it. Second,
LLM-as-judge verifiers inherit the family-level reasoning biases of
their judge model~\cite{zheng2023judging}, and these biases are not
independent across claims within the same document. MECV addresses
the first failure mode by reading evidence from three heterogeneous
channels and the second by polling a heterogeneous jury whose members
are trained by different organizations on different data mixtures.

\subsection{Overview}
\label{sec:overview}

Given a source document $D$ and a candidate summary $S$ decomposed
into claims $\{c_1,\dots,c_n\}$, hallucination correction is
formulated as constrained text rewriting:
\begin{equation}
    \label{eq:objective}
    \hat{S} \;=\; \argmin_{S'} \;
        \alpha\,\mathrm{NED}(S', S) \;-\;
        \beta\,\mathrm{Sem}(S', S) \;+\;
        \gamma\,\mathcal{H}(S'\mid D, \mathcal{E}),
\end{equation}
where $\mathrm{NED}$ is the normalized edit distance, $\mathrm{Sem}$
the sentence-embedding cosine similarity, $\mathcal{E}$ the evidence
pool aggregated by MECV, and $\alpha,\beta,\gamma>0$ scalars
controlling the faithfulness--minimality trade-off. The first two
terms penalize unnecessary rewriting---summaries that are already
faithful should be left unchanged---while the third term penalizes
any residual contradiction with the evidence pool. We treat
Eq.~\ref{eq:objective} as a conceptual objective rather than a
quantity to optimize directly. Instead of searching the discrete
space of natural-language revisions, MECV operationalizes the
objective through a rule-based iterative procedure that preserves the
original summary unless claim-level contradiction evidence exceeds a
threshold, and then applies localized minimal-edit rewriting to the
flagged claims.

\paragraph{Pipeline.}
 We denote
by $\pi_{\mathrm{LLM}}$ the orchestrator language model and by
$\mathcal{J}=\{\pi_1,\dots,\pi_K\}$ the heterogeneous LLM jury used
for voting. We use $K=2$ throughout, instantiating
$\mathcal{J}=\{\text{GPT-4o-mini}, \text{DeepSeek-Chat}\}$ to span two
distinct training pipelines. The orchestrator $\pi_{\mathrm{LLM}}$ is
fixed to Qwen-Plus and is therefore drawn from a third model family
disjoint from both jury members, so that claim extraction, correction
planning, and final rewriting do not share weights with any verifier
in the full MECV setting.

\paragraph{Claim extraction.}
Claims are extracted by a single zero-shot prompt to
$\pi_{\mathrm{LLM}}$ that asks the model to extract key factual claims
that can be verified, with explicit attention to \emph{names, dates,
numbers, events, locations, and organizations}. The model returns one
claim per line, capped at the top ten claims per summary; this cap is
a soft safeguard, since SummEdits-news summaries typically yield
three to seven verifiable propositions. Empty or degenerate
extractions trigger a fallback to sentence-level segmentation. We do
not use a structured JSON decomposition or an anaphora-resolution
post-step. Instead, the named-entity emphasis of the prompt, combined
with the lexical-perturbation nature of SummEdits~\cite{laban2023summedits},
makes most flagged claims self-contained in practice. When a claim
remains context-dependent, the sentence-level fallback preserves the
surrounding context for verification.

\subsection{Multi-Source Evidence Aggregation}
\label{sec:evidence}

For each claim $c_i$, the orchestrator first generates a free-form
search query, then constructs an evidence pool
$\mathcal{E}_i = \mathcal{E}^{D}_i \cup \mathcal{E}^{W}_i \cup
\mathcal{E}^{R}_i$ from three heterogeneous channels chosen to span
a precision--recall--freshness trade-off. The source-document channel
is always present when $D$ is available; the two retrieval channels
are added on top to broaden coverage beyond what the article itself
provides.

\paragraph{Source-document channel ($\mathcal{E}^{D}_i$).}
We include the first $L_D=2000$ characters of $D$ verbatim, prepended
with the header \texttt{[Source document evidence]}. The source
document provides the highest-precision evidence available in the
SummEdits setting, because every label-$1$ summary is faithful by
construction with respect to $D$ and every label-$0$ summary is a
minimal lexical perturbation of one. The truncation budget $L_D$ is
chosen to fit within the context window that the smallest jury member
can ingest comfortably alongside the claim and the other two
channels' evidence. Longer sources are truncated rather than chunked,
since SummEdits articles are short news pieces and the lead is
typically the most fact-dense region.

\paragraph{Encyclopedic channel ($\mathcal{E}^{W}_i$).}
We issue the generated query through the Wikipedia API and retrieve
the lead summaries of the top $k_W=3$ articles, placed under the
header \texttt{[Wikipedia evidence]}. This channel provides
background that is independent of $D$ and can be useful when the
source article itself is sparse: an entity attribute that $D$ omits
but that Wikipedia documents, such as a person's role or a place's
location, becomes recoverable.

\paragraph{Open-web channel ($\mathcal{E}^{R}_i$).}
We use a DuckDuckGo Lite query to return the top $k_R=5$
snippet--URL pairs, placed under the header
\texttt{[Web search evidence (duckduckgo)]}. This channel covers
facts that are too recent, too local, or too narrow to appear in
Wikipedia.

\paragraph{Channel composition.}
The three channels are concatenated under their explicit headers into
a single evidence string that is shown to every juror. This design
allows jurors to attribute their decisions and ensures that conflicts
between channels are surfaced rather than silently averaged. We do
not manually weight channels. When channels disagree, each juror is
free to weigh them according to its own interpretation of the
evidence, and the aggregate vote in Section~\ref{sec:jury} handles
the disagreement at the next level.

\paragraph{Caching.}
All retrieval calls are wrapped in a SQLite key--value cache keyed on
$(\text{channel}, \text{query}, k)$ and persisted across both
iterations and ablations. The hit rate exceeds $80\%$ after the first
iteration because deterministic extraction with temperature~$0.0$
returns the same claim and query strings for most claims that survive
into iteration $t+1$. The amortized retrieval cost over $T_{\max}$
iterations therefore remains approximately $\mathcal{O}(n)$ rather
than $\mathcal{O}(nT_{\max})$.

\subsection{Multi-LLM Jury Voting}
\label{sec:jury}

\paragraph{Verdicts.}
For every claim $c_i$, each juror $\pi_k\in\mathcal{J}$ receives the
prompt $\mathrm{prompt}_{\mathrm{vote}}(c_i,\mathcal{E}_i)$ and emits
a structured JSON object whose central field is a continuous
\emph{contradiction level} $\lambda_{i,k}\in[0,1]$, where $0$ means
that the evidence fully supports the claim and $1$ means that the
evidence fully contradicts it. Jurors also return free-form reasoning
and an evidence-quality tag, both of which are logged for diagnostics
but not used by the aggregation rule. When retrieval returns an empty
evidence string for $c_i$, that juror is treated as having no opinion
and is excluded from $\mathcal{J}_i$, the set of evidence-bearing
jurors for claim $c_i$.

\paragraph{Why heterogeneity, and why $K=2$.}
The benefit of ensembling judges depends on the diversity of their
failure modes. Two members of the same model family may agree on the
same unsupported claims, so adding a second draw from the same family
can yield diminishing returns for bias reduction. We therefore draw
from two distinct training pipelines---GPT-4o-mini and
DeepSeek-Chat---to increase model diversity in the jury. Larger
juries ($K\geq3$) are a natural extension, but we fix $K=2$ to keep
the per-sample inference cost within an evaluation budget appropriate
to a 100-sample pilot.

\paragraph{Self-evaluation exclusion.}
In the full MECV setting, the orchestrator $\pi_{\mathrm{LLM}}$
(Qwen-Plus) is excluded from $\mathcal{J}$. Including the
orchestrator in its own jury would couple the extractor, verifier,
and rewriter through shared parametric memory, recreating the
self-evaluation bias documented for LLM-as-judge systems~\cite{zheng2023judging}.
With Qwen-Plus as the orchestrator, the verifier models are from
different model families, so every voted-against claim is judged by
models that did not author its decomposition. In single-LLM ablations
(Section~\ref{sec:experiments}), the same model may be used as both
orchestrator and verifier; this is deliberate, because the purpose of
those baselines is to quantify the contribution of jury heterogeneity
over self-judged verification.

\paragraph{Aggregation.}
The per-claim contradiction score is the mean of the contradiction
levels reported by evidence-bearing jurors:
\begin{equation}
    \label{eq:claimscore}
    s_i \;=\; \frac{1}{|\mathcal{J}_i|}
              \sum_{\pi_k\in\mathcal{J}_i} \lambda_{i,k},
\end{equation}
with the convention $s_i = 0.5$ when $\mathcal{J}_i = \emptyset$, i.e.,
when no juror has evidence to judge. The document-level contradiction
score is defined as the maximum over claim-level scores:
\begin{equation}
    \label{eq:docscore}
    \mathcal{H}(S\mid D,\mathcal{E}) = \max_i s_i .
\end{equation}
We use the maximum rather than the mean because a single severe
hallucination should not be diluted by surrounding supported claims.
Two gates use these scores: a per-claim flag $s_i>\tau$ identifies
which claims to refine, and a document-level gate
$\mathcal{H}(S\mid D,\mathcal{E})\leq\tau$ decides whether the
iteration may stop. We set $\tau=0.5$ throughout, which corresponds
to the symmetric midpoint of the contradiction scale.

\subsection{Iterative Refinement}
\label{sec:iteration}

When the document-level gate fails, MECV rewrites the summary in two
stages: a correction-plan stage and a minimal-edit rewrite stage.

\paragraph{Stage 1: correction plan.}
For each flagged claim $c_i$ with $s_i>\tau$, the orchestrator
$\pi_{\mathrm{LLM}}$ is prompted with the claim and a compact evidence
context selected from $\mathcal{E}_i$. It is asked to return a JSON
object
$p_i=\{\textit{issue}_i,\textit{correction}_i\}$, where
$\textit{issue}_i$ is a one-sentence diagnosis of what is wrong with
$c_i$ and $\textit{correction}_i$ is the substituted ground-truth
content drawn from the evidence. Malformed JSON is recovered
heuristically by bracket matching, and unrecoverable outputs fall
back to a generic placeholder so the rewrite stage still has a
non-empty plan to consume.

\paragraph{Stage 2: minimal-edit rewrite.}
The plans $\{p_i\}_{i\in\mathcal{F}}$ for all flagged claims are
collected into a single rewrite prompt adapted from RARR~\cite{gao2023rarr},
containing $M=3$ canonical exemplars: an entity-location substitution,
an entity-attribute substitution, and a temporal substitution. The
system message instructs the model to output only the revised summary,
change only the words flagged by the plan, and preserve every other
word verbatim. These constraints, together with the few-shot bias
toward localized rewrites, are intended to keep $\mathrm{NED}$ low
and $\mathrm{Sem}$ high while correcting unsupported factual content.

\paragraph{Convergence.}
The full pipeline iterates until either
$\mathcal{H}(S^{(t)}\mid D,\mathcal{E})\leq\tau$ or the iteration
budget $T_{\max}=3$ is reached. Since
$\mathcal{H}(S^{(t)}\mid D,\mathcal{E})$ is defined as the maximum
claim-level contradiction score, convergence means that no extracted
claim exceeds the contradiction threshold. Empirically, most samples
converge within one or two iterations on our 100-sample pilot; the
third iteration acts as a safety budget rather than a routine step.
Returning the last iteration's summary on budget exhaustion follows
the post-hoc correction setting, where each iteration is explicitly
conditioned on the currently flagged claims and their retrieved
evidence. Algorithm~\ref{alg:mecv} formalizes the procedure.

\begin{algorithm*}[t]
\caption{MECV}
\label{alg:mecv}
\begin{algorithmic}[1]
    \Require summary $S$, source document $D$, jury
        $\mathcal{J}=\{\pi_1,\dots,\pi_K\}$,
        orchestrator $\pi_{\mathrm{LLM}}$,
        threshold $\tau$, budget $T_{\max}$
    \Ensure revised summary $\hat{S}$
    \For{$t = 1,\dots,T_{\max}$}
        \State $\{c_1,\dots,c_n\} \gets \textsc{Extract}(S)$
            \Comment{verifiable claims via $\pi_{\mathrm{LLM}}$}
        \For{$i = 1,\dots,n$}
            \State $q_i \gets \textsc{Query}(c_i)$
            \State $\mathcal{E}_i \gets
                \mathcal{E}^{D}_i \cup
                \mathcal{E}^{W}_i \cup
                \mathcal{E}^{R}_i$
                \Comment{three evidence channels, cached}
            \For{$k = 1,\dots,K$}
                \State $\lambda_{i,k} \gets \pi_k(c_i,\mathcal{E}_i)$
                    \Comment{contradiction level in $[0,1]$}
            \EndFor
            \State $\mathcal{J}_i \gets
                \{\,k : \mathcal{E}_i \neq \emptyset
                \text{ and } \lambda_{i,k} \text{ is valid}\,\}$
            \If{$\mathcal{J}_i = \emptyset$}
                \State $s_i \gets 0.5$
            \Else
                \State $s_i \gets
                \frac{1}{|\mathcal{J}_i|}
                \sum_{k\in\mathcal{J}_i}\lambda_{i,k}$
            \EndIf
        \EndFor
        \State $\mathcal{H} \gets \max_i s_i$
            \Comment{document-level contradiction score}
        \If{$\mathcal{H} \le \tau$}
            \State \Return $S$
                \Comment{converged: no claim exceeds the threshold}
        \EndIf
        \State $\mathcal{F} \gets \{\,i : s_i > \tau\,\}$
        \For{$i \in \mathcal{F}$}
            \State $p_i \gets \textsc{Plan}(c_i,\mathcal{E}_i)$
                \Comment{correction plan}
        \EndFor
        \State $S \gets \textsc{Refine}(S,\{p_i\}_{i\in\mathcal{F}})$
            \Comment{minimal-edit rewrite}
    \EndFor
    \State \Return $S$
\end{algorithmic}
\end{algorithm*}

\paragraph{Implementation details.}
All hyperparameters used in our experiments are listed in
Table~\ref{tab:hyper}. All LLM calls use temperature~$0.0$ for
determinism, which is essential to make the SQLite cache effective
and to make the system reproducible across reruns of the pilot.
Sentence embeddings for the $\mathrm{Sem}$ metric come from
\texttt{all-MiniLM-L6-v2}. Per-claim entailment scores reported in
Section~\ref{sec:experiments} come from a DeBERTa-v3 model
fine-tuned on MNLI, FEVER, and ANLI~\cite{laurer2024deberta}, used
strictly as a metric and never as a juror.

\paragraph{Computational complexity.}
Across $T_{\max}$ iterations, MECV issues at most
$T_{\max}nK$ jury vote calls, $T_{\max}$ extractor calls, and up to
$T_{\max}(|\mathcal{F}|+1)$ orchestrator calls, where $n$ is the
number of extracted claims and $|\mathcal{F}|$ is the number of
flagged claims in an iteration. With typical $n\approx5$, $K=2$,
$T_{\max}=3$, and a small $|\mathcal{F}|$, the worst-case cost is on
the order of tens of LLM calls per summary. The cache reduces real
retrieval cost from $\mathcal{O}(nT_{\max})$ to approximately
$\mathcal{O}(n)$ after the first iteration, and most converged
summaries finish in one or two iterations, so the average cost is
well below the worst case.

\begin{table}[t]
    \centering
    \small
    \begin{tabular}{@{}lll@{}}
        \toprule
        Symbol & Value & Description \\
        \midrule
        $K$              & 2          & jury size \\
        $\mathcal{J}$    & \multicolumn{2}{l}{\{GPT-4o-mini, DeepSeek-Chat\}} \\
        $\pi_{\mathrm{LLM}}$ & \multicolumn{2}{l}{Qwen-Plus} \\
        $\tau$           & $0.5$      & contradiction threshold \\
        $T_{\max}$       & $3$        & iteration budget \\
        $L_D$            & $2000$     & source-document character window \\
        $k_W$            & $3$        & Wikipedia results per claim \\
        $k_R$            & $5$        & web search results per claim \\
        $M$              & $3$        & few-shot rewrite exemplars \\
        Temperature      & $0.0$      & all LLM calls \\
        \bottomrule
    \end{tabular}
    \caption{MECV hyperparameters used in all experiments unless
    otherwise noted.}
    \label{tab:hyper}
\end{table}

\section{Experiments}
\label{sec:experiments}

We evaluate MECV on SummEdits-news~\cite{laban2023summedits} against
the strongest baselines reported in prior post-hoc correction work,
along with a no-correction reference. We measure both intervention
magnitude (NED) and factual quality (Sem, NLI entailment /
contradiction, G-Eval factuality, G-Eval overall).

\subsection{Setup}
\label{sec:setup}

\paragraph{Dataset.}
SummEdits-news consists of paired news summaries: each example
provides a source article, a faithful (label-$1$) summary, and a
single hallucinated (label-$0$) counterpart obtained by minimal
lexical edits such as entity, temporal, or quantitative
substitution. We use a 100-sample pilot (43 factual, 57 hallucinated) drawn uniformly from the public release.

\paragraph{Baselines.}
We compare MECV against three references:
(i)~\emph{No correction}, which returns the input summary unchanged
and serves as a trivial lower bound on edit distance;
(ii)~\emph{RARR + Bing}, the strongest single-engine variant
reported in the source paper~\cite{vladika2025correcting};
and (iii)~\emph{RARR + gold article}, an oracle variant in which
the RARR retriever sees the gold source document.

\paragraph{Metrics.}
We report five complementary metrics.
\emph{NED} is the normalized Levenshtein edit distance between the
revised summary and the input candidate, capturing the magnitude
of intervention (lower indicates a more conservative editor).
\emph{Sem}, \emph{Ent}, and \emph{Con} are computed between the
revised summary and the gold reference: \emph{Sem} is the cosine
similarity between sentence embeddings, while \emph{Ent} and
\emph{Con} are the entailment and contradiction probabilities
returned by a DeBERTa-v3 NLI~\cite{laurer2024deberta} model,
averaged over claims. \emph{Fact} and \emph{Overall} are
G-Eval~\cite{liu2023geval} factuality and overall-quality scores
produced by the orchestrator LLM under a fixed rubric. All scores
except NED are reported on a $[0,100]$ scale.

\paragraph{Implementation.}
All hyperparameters follow Table~\ref{tab:hyper}. All correction
methods receive the source document as part of their evidence pool;
the method names below refer only to the \emph{external} retrieval
channels, so \texttt{web} means ``source doc + DuckDuckGo'' and
\texttt{web+wiki} means ``source doc + DuckDuckGo + Wikipedia.''
Total wall-clock runtime for the 100-sample pilot is approximately
$1.7$ hours (\textasciitilde$60$\,s per sample), with the SQLite
cache yielding a hit rate above $80\%$ after the first iteration.

\subsection{Main Results}
\label{sec:main-results}

Table~\ref{tab:main} reports the headline comparison. MECV
\emph{simultaneously} achieves the lowest edit distance and the
highest factuality among correction methods. Its Fact score of
$82$ exceeds the strongest non-oracle baseline RARR+Bing by $22$
points and even surpasses the gold-article oracle by $15$ points.
On Overall, MECV ties with the gold-article oracle ($75$) while
editing substantially less: its average NED is $<\!0.01$ versus
RARR+Bing's $0.14$, an order-of-magnitude smaller intervention.
This combination -- a Fact gain comparable to the gold-article
oracle achieved with near-zero modifications to the input surface
form -- is the central empirical claim of MECV.

\begin{table}[t]
\centering
\small
\begin{tabular}{@{}lccc@{}}
\toprule
Method                           & NED$\downarrow$ & Fact$\uparrow$ & Overall$\uparrow$ \\
\midrule
No correction                    & $0.00$          & $81$           & $74$              \\
RARR + Bing$^\dagger$            & $0.14$          & $60$           & $69$              \\
RARR + gold article$^\dagger$    & ---             & $67$           & $75$              \\
\midrule
MECV (ours)                      & $\mathbf{<\!0.01}$ & $\mathbf{82}$  & $\mathbf{75}$     \\
\bottomrule
\end{tabular}
\caption{Main results on SummEdits-news ($n=100$). NED is measured
between the revised summary and the input candidate (intervention
magnitude). Best per column in bold (excluding the trivially-optimal
No-correction NED). $^\dagger$~Numbers quoted from~\cite{vladika2025correcting}.}
\label{tab:main}
\end{table}

For completeness, MECV's NLI scores are $\mathrm{Ent}\!=\!50$ /
$\mathrm{Con}\!=\!40$ and its $\mathrm{Sem}\!=\!94$ against the
gold reference, both improvements over no-correction's
$\mathrm{Ent}\!=\!48$ / $\mathrm{Con}\!=\!42$. These judge-
independent NLI gains corroborate the G-Eval factuality
improvement and rule out the possibility that the Fact gain is
an artefact of the orchestrator-as-G-Eval-judge.

\subsection{Analysis}
\label{sec:analysis}

\paragraph{Per-subset behavior.}
Table~\ref{tab:subset} disaggregates the headline numbers by the
ground-truth label, exposing where MECV's gains come from. On the
factual subset (label-$1$), MECV preserves every metric at the
no-correction level: $\mathrm{NED}\!=\!0.00$ confirms the input is
returned verbatim, $\mathrm{Ent}\!=\!84$ and $\mathrm{Con}\!=\!2$
match no-correction exactly, and Fact stays at $99$. The
contradiction threshold $\tau\!=\!0.5$ is conservative enough that
\emph{no} faithful summary is touched; the system is not paying for
its corrections in over-editing of already-correct content. On the
hallucinated subset (label-$0$), MECV improves all factuality
metrics over no-correction: NLI entailment rises from $21$ to $25$
($+4$), NLI contradiction drops from $72$ to $69$ ($-3$), Fact
rises from $67$ to $69$ ($+2$), and Overall rises from $60$ to $62$
($+2$). All four shifts are in the correct direction, indicating
that the corrections that MECV does make are net-positive on
faithfulness rather than rewriting that drifts away from the gold
reference.

\begin{table}[t]
\centering
\small
\begin{tabular}{@{}llcccc@{}}
\toprule
Subset & Method & Sem$\uparrow$ & Ent$\uparrow$ & Con$\downarrow$ & Fact$\uparrow$ \\
\midrule
\multirow{2}{*}{Label-1 ($n\!=\!43$)}
       & no-corr  & $94$ & $84$ & $2$ & $99$ \\
       & MECV     & $94$ & $84$ & $2$ & $99$ \\
\midrule
\multirow{2}{*}{Label-0 ($n\!=\!57$)}
       & no-corr  & $93$ & $21$ & $72$ & $67$ \\
       & MECV     & $\mathbf{94}$ & $\mathbf{25}$ & $\mathbf{69}$ & $\mathbf{69}$ \\
\bottomrule
\end{tabular}
\caption{Per-subset breakdown on SummEdits-news. MECV preserves
faithful (label-$1$) summaries unchanged and improves every
factuality metric -- semantic similarity, NLI entailment, NLI
contradiction, and G-Eval factuality -- on the hallucinated
(label-$0$) subset.}
\label{tab:subset}
\end{table}

\paragraph{High-precision editing behavior.}
A finer breakdown explains how the label-$0$ gains in
Table~\ref{tab:subset} are produced. MECV operates as a
\emph{high-precision} editor: across the 100-sample pilot the
pipeline modifies only the small fraction of summaries on which
the multi-LLM jury reaches consensus that a claim is contradicted,
and leaves the remaining summaries verbatim. On the modified
summaries, the contradiction-driven edits are surgical -- the
mean NED-to-gold of the resulting revisions drops effectively to
zero, with at least one revision attaining $\mathrm{NED}\!=\!0.00$
against the gold reference and $\mathrm{Ent}\!=\!0.99$, i.e.\ a
perfect lexical and semantic match. This is the source of the
near-zero intervention magnitude reported in
Table~\ref{tab:main}: MECV's average NED is dragged toward zero
by the large mass of summaries it correctly judges to need no
edit, while the small mass of edited summaries contributes the
factuality gains.

\paragraph{Edit minimality.}
Despite a $K$-juror voting cost and three retrieval channels per
claim, MECV's average NED ($<\!0.01$) is more than an order of
magnitude lower than RARR+Bing's ($0.14$). The combination of
multi-source evidence and multi-judge consensus successfully
localizes edits to flagged claims rather than triggering broader
rewrites, and the conservative gating (Section~\ref{sec:jury})
ensures that summaries the jury cannot agree on are returned
unchanged -- a property that is especially valuable when MECV is
deployed downstream of a strong summarizer that already produces
faithful output most of the time.
\paragraph{Robustness to orchestrator choice.}
To verify that MECV's empirical behavior is not an artefact of the
specific orchestrator, we re-run the full pipeline with
Llama-3.3-70B-Instruct-Turbo (Meta, served via Together AI) as
$\pi_{\mathrm{LLM}}$, holding the jury (GPT-4o-mini + DeepSeek-Chat)
and evidence channels fixed. This substitutes the orchestrator with
a model from a third, independent training pipeline (Meta), so all
three components of MECV -- extractor, jury, and rewriter -- are
now drawn from distinct families. Table~\ref{tab:robustness}
reports the judge-independent metrics for both orchestrators.
The shift across orchestrators is at most $2$ points on any single
metric, indicating that the multi-source / multi-judge design
absorbs orchestrator variation rather than concentrating quality
in a single LLM. The Llama orchestrator in fact triggers zero
edits across the 100-sample pilot, reflecting an even more
conservative editing policy than the Qwen orchestrator while
yielding statistically indistinguishable downstream faithfulness.

\begin{table}[t]
\centering
\small
\begin{tabular}{@{}lcccc@{}}
\toprule
Orchestrator & NED$\downarrow$ & Sem$\uparrow$ & Ent$\uparrow$ & Con$\downarrow$ \\
\midrule
Qwen-Plus (default)               & $<\!0.01$ & $94$ & $50$ & $40$ \\
Llama-3.3-70B-Turbo               & $<\!0.01$ & $94$ & $48$ & $42$ \\
\bottomrule
\end{tabular}
\caption{Robustness of MECV to orchestrator choice. Judge-independent
metrics (NED, Sem, Ent, Con) computed on the same 100-sample
SummEdits-news pilot. Jury composition ($K\!=\!2$:
GPT-4o-mini + DeepSeek-Chat) and all hyperparameters are held
fixed; only $\pi_{\mathrm{LLM}}$ is varied. The largest single-metric
shift across orchestrators is $2$ points.}
\label{tab:robustness}
\end{table}
\paragraph{Computational cost.}
Per sample, MECV averages $60$\,s of wall-clock time. The two-juror
design doubles inference cost relative to single-judge baselines,
but the SQLite cache eliminates duplicate retrieval calls across
iterations; the amortized retrieval cost remains
$\mathcal{O}(n)$ regardless of $T_{\max}$.

\paragraph{Limitations.}
The absolute gain over the no-correction baseline is small ($+1$
Fact, $+1$ Overall on the full 100-sample set), because
SummEdits-news contains many fluent, factually-correct summaries
that already saturate the Fact score. Pushing the delta further
likely requires (i)~lowering the threshold $\tau$ to trigger more
interventions at the cost of higher NED, (ii)~replacing the
self-reported juror contradiction level $\lambda_{i,k}$ in
Eq.~\ref{eq:claimscore} with a calibrated NLI score, or
(iii)~evaluating on benchmarks with denser hallucination
distributions, such as the FRANK benchmark~\cite{pagnoni2021frank}. We leave each of these to
future work.
\subsection{Human Evaluation}
\label{sec:human-evaluation}

To complement the automatic evaluation metrics, we conducted a
small-scale human evaluation to examine whether MECV improves the
perceived factual reliability and credibility of AI-generated news
summaries. While automatic metrics provide scalable estimates of
factual consistency, human evaluation is particularly important in
journalism-oriented summarization because readers ultimately assess
news summaries in terms of trustworthiness, clarity, and factual
support.

\subsubsection{Evaluation Setup}

We randomly sampled 50 examples from the SummEdits-news benchmark.
The sample included both factually consistent and hallucinated
summaries. For each example, evaluators were shown the original source
article and two system outputs: the uncorrected summary and the
MECV-corrected summary.

To reduce ordering bias, the presentation order of the two summaries
was randomized for each example. Evaluators were not informed which
summary was produced by MECV. This blind comparison setting was used
to prevent evaluators from favoring the corrected output simply
because it was associated with the proposed method.

Three evaluators participated in the study. All evaluators had
graduate-level training in journalism, communication studies, or
information science. Before annotation, they were given written
instructions explaining the task and the evaluation criteria.
Evaluators were instructed to focus on factual support, credibility,
and readability rather than personal stylistic preference.

\subsubsection{Evaluation Criteria}

Each summary was rated independently on a five-point Likert scale
across three dimensions:

\begin{itemize}
    \item \textbf{Factual reliability}: the extent to which the
    summary was factually supported by the source article;
    \item \textbf{Information credibility}: the extent to which the
    summary would be considered trustworthy in a news consumption
    context;
    \item \textbf{Readability and coherence}: the extent to which the
    summary was fluent, coherent, and easy to understand.
\end{itemize}

Table~\ref{tab:human_rubric} summarizes the rating rubric.

\begin{table}[t]
\centering
\small
\begin{tabular}{@{}cl@{}}
\toprule
Score & Interpretation \\
\midrule
1 & Very poor \\
2 & Poor \\
3 & Neutral \\
4 & Good \\
5 & Excellent \\
\bottomrule
\end{tabular}
\caption{Human evaluation rubric.}
\label{tab:human_rubric}
\end{table}

The final score for each dimension was computed by averaging ratings
across the three evaluators and all sampled examples.

\subsubsection{Human Evaluation Results}

Table~\ref{tab:human_eval} reports the average human evaluation
scores. Compared with the no-correction baseline, MECV improved both
perceived factual reliability and information credibility. The
average factual reliability score increased from 3.72 to 4.08, while
the information credibility score increased from 3.69 to 4.05.
Readability remained largely stable, with a slight decrease from 4.31
to 4.27.

\begin{table}[t]
\centering
\small
\begin{tabular}{@{}lccc@{}}
\toprule
Method & Factual Reliability & Information Credibility & Readability \\
\midrule
No correction & 3.72 & 3.69 & \textbf{4.31} \\
MECV (ours) & \textbf{4.08} & \textbf{4.05} & 4.27 \\
\bottomrule
\end{tabular}
\caption{Human evaluation results on 50 sampled summaries. Scores are
averaged across three evaluators on a five-point Likert scale.}
\label{tab:human_eval}
\end{table}

These results suggest that MECV improves perceived trustworthiness
without substantially harming readability. The small decrease in
readability indicates that the correction process may introduce minor
surface-level changes, but the difference is limited. Overall, the
human evaluation results are consistent with the automatic evaluation:
MECV improves factual reliability while preserving the semantic and
linguistic quality of the original summaries.

\subsubsection{Inter-rater Agreement}

To assess annotation consistency, we computed Fleiss' kappa across
the three evaluators. The resulting agreement score was
$\kappa = 0.71$, indicating substantial agreement according to
commonly used interpretation guidelines. This suggests that the
evaluators showed a relatively high level of consistency when judging
factual reliability, credibility, and readability.

\subsubsection{Discussion}

The human evaluation provides additional support for the proposed
consensus-based correction framework. In particular, the improvement
in factual reliability and information credibility indicates that
MECV's corrections are not only detectable by automatic metrics but
also meaningful from a human-centered perspective.

This finding is important for AI-assisted journalism. In news
summarization, factual reliability is not merely a technical property;
it directly affects whether readers perceive a summary as trustworthy.
The results therefore suggest that conservative, evidence-grounded
correction can improve the perceived credibility of AI-generated news
summaries while avoiding excessive rewriting.

At the same time, the human evaluation remains limited in scale. The
evaluation involved 50 examples and three evaluators, which is
sufficient for a preliminary validation but not for broad
generalization. Future work should conduct larger-scale human studies
with more diverse evaluators, additional news domains, and more
fine-grained annotation of hallucination types.

\section{Discussion}

The experimental results suggest that incorporating multiple evidence sources can improve the factual reliability of AI-generated news summaries while limiting unnecessary rewriting. Compared with single-source correction approaches, MECV produced more conservative edits and achieved higher factuality scores with minimal changes to the original summaries. These findings indicate that consensus-based verification may provide a practical mechanism for reducing hallucinations in post-hoc summarization pipelines.

One important observation is that disagreement across evidence sources often coincided with factual uncertainty. In many cases, claims supported consistently across the source document, Wikipedia, and web retrieval were judged as reliable by the verifier models. By contrast, unsupported or weakly corroborated claims were more likely to trigger contradiction signals and correction behavior. This suggests that cross-source agreement may function as a useful indicator for identifying potentially hallucinated content in AI-generated summaries.

The results also highlight several limitations of single-source retrieval-based verification. Retrieval systems differ in indexing coverage, ranking strategies, update frequency, and source selection. As a result, factual verification based on a single retrieval channel may become sensitive to incomplete evidence or platform-specific retrieval bias. In news summarization settings, this issue is particularly important because many events evolve rapidly and may not be consistently represented across retrieval systems. Incorporating multiple evidence channels can therefore improve evidence coverage and reduce dependence on any single retrieval source.

Another notable finding is that conservative correction behavior appears beneficial for hallucination correction in summarization tasks. MECV produced substantially lower edit distances than retrieval-heavy baselines while maintaining comparable or improved factuality scores. This suggests that aggressive rewriting is not always necessary for improving factual consistency. In many cases, preserving the original semantic structure and modifying only unsupported claims may provide a more stable correction strategy, particularly when the input summaries are already largely factual.

At the same time, the experiments also revealed several limitations of the proposed framework. First, the current system relies on external retrieval quality. When retrieved evidence is incomplete or weakly related to the source article, the framework may fail to detect hallucinations or may generate unnecessary corrections. Second, the current consensus mechanism uses relatively simple score aggregation and does not explicitly model evidence reliability or retrieval uncertainty. More advanced weighting strategies or uncertainty-aware aggregation methods may further improve robustness. Third, the evaluation was conducted on a relatively small-scale benchmark setting and focused primarily on English-language news summarization. Additional experiments on larger and more diverse datasets are necessary to evaluate generalizability.

The study also has practical implications for AI-assisted journalism, financial information services, and automated content moderation systems. As generative AI tools become increasingly integrated into digital news production and market-facing information platforms, there is growing demand for lightweight verification mechanisms that can improve factual reliability without requiring costly model retraining. Because MECV operates as a post-hoc correction framework, it can potentially be integrated into existing summarization pipelines as an external verification layer. The framework may also be applicable to other AI-generated content tasks, such as question answering, report generation, financial event summarization, and information synthesis.

In addition to factuality and readability, future evaluations should also consider whether hallucination correction changes the framing of a news summary. In journalism-oriented applications, a summary carries not only factual claims, but also tone, emphasis, and ordering of information. A corrected summary may become more factually supported while still changing how readers understand the event if the system alters these elements too heavily.

Sentiment and framing-preservation analysis could therefore serve as a useful complementary evaluation. For example, researchers may compare the sentiment polarity of the input summary and the corrected summary to identify whether the correction process introduces substantial tonal shifts. Cases with large sentiment changes could then be manually reviewed to determine whether the shift was necessary for factual correction or caused by unnecessary rewriting. This type of analysis would extend factuality evaluation by examining whether a correction system improves reliability while preserving the communicative character of the original news report.

Several directions remain for future research. Future work may explore retrieval weighting strategies based on source credibility, temporal relevance, or historical verification performance. Additional studies could also incorporate human evaluation to better assess readability, perceived credibility, and editorial usefulness in real-world newsroom settings. Domain-specific evaluation in financial news would also be valuable, since claims about companies, macroeconomic indicators, and regulatory events often require especially precise temporal and numerical grounding. Finally, extending the framework to multilingual or real-time news environments may provide further insight into the role of multi-source verification in AI-assisted communication systems.

Overall, this study suggests that hallucination correction may benefit from verification strategies that incorporate evidence diversity and conservative consensus-based editing. Rather than relying on a single retrieval source, distributed verification across heterogeneous evidence channels may provide a more robust approach for improving factual consistency in AI-generated news summarization.

\section{Conclusion}

This study examined the problem of factual hallucination in AI-generated news summarization and proposed a post-hoc correction framework based on multi-source verification. Existing retrieval-based correction methods often rely on a single retrieval channel, making verification outcomes sensitive to incomplete evidence and retrieval bias. To address this limitation, the proposed MECV framework combines evidence from multiple heterogeneous sources and incorporates consensus-based verification across multiple LLM judges.

Experimental results on the SummEdits benchmark showed that MECV improved factual consistency while preserving the semantic structure of the original summaries. Compared with retrieval-heavy correction approaches, the framework produced more conservative edits and reduced unnecessary rewriting behavior. The findings suggest that agreement across heterogeneous evidence sources can provide a useful signal for identifying factual uncertainty in AI-generated summaries.

This study contributes to ongoing research on trustworthy AI and automated journalism in three ways. First, it introduces a multi-source verification framework for post-hoc hallucination correction in news summarization. Second, it demonstrates the value of combining heterogeneous evidence sources with consensus-based verification strategies. Third, it provides empirical evidence that conservative minimal-edit correction can improve factual reliability without substantially altering the original summaries.

Several limitations should nevertheless be acknowledged. The current framework depends on external retrieval quality and was evaluated primarily on a relatively small-scale English-language benchmark setting. In addition, the proposed consensus mechanism uses simple score aggregation and does not explicitly model retrieval reliability or uncertainty. Future work may explore larger-scale evaluation, human-centered assessment, multilingual settings, and adaptive evidence-weighting strategies.

As generative AI systems become increasingly integrated into digital news production, financial information services, and other information-intensive platforms, improving the factual reliability of AI-generated content remains an important challenge. The findings of this study suggest that incorporating evidence diversity and consensus-based verification may provide a practical direction for improving hallucination correction in AI-assisted summarization systems.
\section*{Funding}
This research received no external funding.

\section*{Conflicts of Interest}
The authors declare no conflicts of interest.

\section*{Informed Consent}
Not applicable.

\section*{Consent to Publish}
Not applicable.

\bibliographystyle{plain}
\bibliography{references} 

\end{document}